\documentclass{dyno2025}

\usepackage{times, soul}  
\usepackage{helvet,nicefrac} 
\usepackage{courier}  
\usepackage[small]{caption} 
\usepackage{subcaption}
\usepackage{mwe}
\usepackage{xcolor,amsmath,amsfonts,amssymb}

\usepackage{todonotes}

\newtheorem{problem}{Problem}
\newcommand{\diag}{\rm{diag}}

\title{The Silence that Speaks: Neural Estimation via Communication Gaps}

\optauthor{%
\Name{Shubham Aggarwal} \Email{sa57@illinois.edu}\\
\addr University of Illinois Urbana Champaign, USA
\AND
\Name{Dipankar Maity} \Email{dmaity@charlotte.edu}\\
\addr University of North Carolina at Charlotte, USA
\AND
\Name{Tamer Ba\c{s}ar} \Email{basar1@illinois.edu}\\
\addr University of Illinois Urbana Champaign, USA}

\begin{document}

\maketitle

\begin{abstract}%
Accurate remote state estimation is a fundamental component of many autonomous and networked dynamical systems, where multiple decision-making agents interact and communicate over shared, bandwidth-constrained channels.
These communication constraints introduce an additional layer of complexity, namely, the decision of \textit{when to communicate}. This results in a fundamental trade-off between estimation accuracy and communication resource usage. Traditional extensions of classical estimation algorithms (e.g., the Kalman filter) treat the absence of communication as `missing' information. 
However, silence itself can carry implicit information about the system’s state, which, if properly interpreted, can enhance the estimation quality even in the absence of explicit communication.
Leveraging this implicit structure, however, poses significant analytical challenges, even in relatively simple systems. 
In this paper, we propose \texttt{CALM} (\underline{C}ommunication-\underline{A}ware \underline{L}earning and \underline{M}onitoring), a novel learning-based framework that jointly addresses the dual challenges of communication scheduling and estimator design. 
Our approach entails learning not only \textit{when to communicate} but also \textit{how to infer useful information from periods of communication silence}. We perform comparative case studies on multiple benchmarks to demonstrate that \texttt{CALM} is able to decode the implicit coordination between the estimator and the scheduler to extract information from the instances of `silence' and enhance the estimation accuracy. 
\end{abstract}

\section{Introduction}
Remote state estimation plays a foundational role in a wide range of engineering applications, including channel state tracking in cellular networks, spectrum sensing in cognitive radio systems, trajectory prediction in autonomous space exploration, and load forecasting in power grids \cite{soderstrom2012discrete,thilina2013machine}, to name a few. In this work, we focus on the problem of \textit{remote estimation of a stochastically evolving discrete-time dynamical system}. The overall system (as shown in Fig.~\ref{Fig:systemmodel}) comprises a dynamically evolving process along with a team of two cooperating decision-making agents: (1) a collocated scheduler and (2) a remote estimator. The scheduler continuously observes the system state and decides \textit{when to communicate} this information to the remote estimator, which in turn produces the best possible estimate of the state. The primary objective is to balance the competing goals of maintaining estimation accuracy and minimizing communication overhead.

Departing from classical estimation frameworks, such as the Kalman filter \cite{kalman1960new} and its numerous variations that treat non-transmissions as missing information, the core contribution of this work is to leverage the \textit{implicit information} contained in communication gaps (i.e., intervals of silence between  transmissions). Our approach is inspired by principles in neuroscience and cognitive systems, where humans can infer intent or internal state even in the absence of sensory input—for example, interpreting pauses or hesitations in speech—making the absence of a signal, a signal in its own right. The key technical challenge in exploiting this insight lies in the fact that the estimation dynamics become tightly coupled with the communication policy, which renders the estimator analytically intractable—even in linear systems for which the Kalman filter was originally developed. Moreover, the scheduler policy is unknown and lacks a closed-form solution.

To address these challenges, we propose \texttt{CALM}, an alternating deep reinforcement learning (DRL) framework based on Proximal Policy Optimization (PPO) that jointly learns both the scheduling policy and a nonlinear estimator. Crucially, the learned estimator is capable of extracting structure from communication silence. Empirical results across several benchmark control domains demonstrate that \texttt{CALM} significantly outperforms traditional linear estimation techniques and heuristic scheduling baselines by effectively exploiting latent information embedded in the no-communication events.

\section{Related Work}

Team decision problems have been extensively studied since the foundational works of Marschak and Radner~\cite{marschak1955elements,radner1962team}, and serve as a cornerstone for the formulation of distributed decision-making problems involving multiple agents. In such settings, decision-makers (DMs) do not have access to common (centralized) information and collaborate to optimize a common objective. Due to the distributed nature of this setup, coupled with the dynamic nature of the underlying Markov Decision Process (MDP), the information structure becomes critical and significantly influences tractability and optimality~\cite{feldbaum1961dual,yuksel2013stochastic}.

The co-design problem of scheduling and estimation considered in this work can be viewed as a special case of a team decision problem involving\textit{ two DMs: the scheduler and the estimator}. 
The joint design of control/estimation and scheduling policies has been explored in several prior works~\cite{imer2010optimal,lipsa2011remote,molin2014optimal,molin2017event,maity2020minimal}. Under a partially nested information structure, it is known that for linear systems, a certainty equivalence property holds, allowing the controller to be designed first (as a function of the conditional estimate of the state), and subsequently the scheduler. However, even in such cases, the conditional estimate may not admit a closed-form expression, making the estimation-scheduling co-design, i.e., the problem in this work, fundamentally hard.

To sidestep analytical intractability, most prior approaches restricted the estimator to be of a linear recursive form \cite{molin2014optimal,eisen2022communication,wang2024infinite}.
This was often justified by introducing assumptions on the estimator's information set—e.g., partially/completely ignoring the scheduling instants' information \cite{molin2014optimal,molin2017event} or by limiting the problem to scalar systems \cite{lipsa2011remote,molin2014optimal}, symmetric policy spaces, or symmetric noise distributions \cite{ramesh2013design,hertneck2025current}. These assumptions allowed the estimator to be designed in closed form for a given scheduling policy. Attempts to address the informational value of no-communication events have been made in literature~\cite{molin2017event,maity2020minimal}, where a preliminary complexity analysis was provided. Recent work~\cite{soleymani2022value} showed that for a stochastic linear system with additive Gaussian noise and one-step delayed communication, silence does not affect estimation, implying that the estimator can remain linear. Nevertheless, the optimal scheduler remains analytically intractable due to the bilinear nature of the estimation error dynamics~\cite{aggarwal2025interq}. The challenge becomes even more pronounced for general nonlinear systems when one attempts to incorporate the full information available at the estimator, including the no-communication instants. Doing so makes the estimator’s dynamics nonlinear and non-recursive, breaking the structure that classical solutions rely on, and this has been largely unaddressed in literature.

\textbf{Contributions: }Motivated by the above key technical challenges, in this work we adopt a DRL perspective to study and solve the aforementioned co-design problem.
 Unlike existing literature, we consider the co-design of the scheduler-estimator policy in remote estimation systems under most general information structures, arbitrary noise distributions and general closed-loop policy classes. To alleviate the consequent intractability yielded within the estimation (and also the scheduling policy) computation, we propose \texttt{CALM}---a DRL-based algorithm for co-designing both the scheduler and the estimator, with the main objective of capturing the latent information present within the periods of silence. 

We emphasize that our setup is different from the classical control-oriented neuro-Lyapunov methods, where the objective is that of \textit{controller design under full state observability} \cite{chang2019neural,zhou2022neural,wu2023neural} rather than estimator design.
Additionally, ours appears to be the first work which deals with general nonlinear stochastic dynamics with arbitrary noise distributions, without restrictions on the noise probability density functions. Finally, our extensive numerical experiments on multiple benchmark control tasks provide insights into how implicit information from silence can be used to infer the underlying structure within the stochastic noise, thereby improving estimation accuracy over existing baselines while keeping communication costs low.

\textbf{Notations:} For an integer $m$, we let $[m]:= \{0, 1, 2, \cdots, m\}$. For square matrices $A$ and $B$, we define $\diag(A,B):= \begin{pmatrix}
    A & 0 \\
    0 & B
\end{pmatrix}$. For a positive semi-definite matrix $\Gamma$ and a vector $x$, we define $\|x\|^2_\Gamma:= x^\top \Gamma x$.

\section{Problem Formulation}
\begin{figure}[t]
	\centering	\includegraphics[width=0.6\columnwidth ]{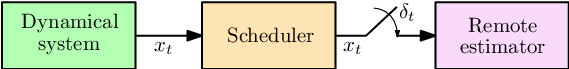}
	\caption{\small{Schematic of a remote estimation system.
	}}
	\label{Fig:systemmodel}
 	\vspace{-0.2cm}
\end{figure}

Consider a stochastic dynamical system evolving as:
\begin{align}\label{eq:nonlinsystem}
    x_{t+1} = f(x_t, w_t), \quad t \ge 0,
\end{align}
where $x_t \in \mathbb R^n$ denotes the state
and $w_t \in \mathbb R^m$ denotes an i.i.d. noise sample drawn from a zero-mean distribution $\mathcal{P}$, which is assumed to have finite second moments.\footnote{The zero-mean assumption is without any loss of generality. If $\mathbb{E}[w_t] = \bar{w}$, we define $\tilde{w}_t = w_t - \bar{w}$ and $x_{t+1} = f(x_t, w_t) = f(x_t, \bar{w}+\tilde{w}_t)  := \tilde{f}(x_t, \tilde{w}_t)$.} The initial state $x_0$ is also sampled from a zero-mean distribution $\mathcal{P}_0$ with finite second moment and is assumed to be independent of the noise distribution $\mathcal{P}$. 

The state $x_t$ is actively communicated by a collocated scheduler/sensor to a remote estimator for data logging and analysis, as shown in Fig. \ref{Fig:systemmodel}. However, due to constraints associated with remote communication (e.g., limited power or bandwidth), transmissions must be scheduled in an \textit{efficient} manner. Naturally, more frequent communication improves estimation accuracy at the expense of higher communication costs, and vice versa. As an illustrative application, consider a Mars rover exploring the Martian surface. It intermittently connects to a NASA base station to transmit its current state. Given strict energy and bandwidth constraints, the question arises: \textit{how frequently should the rover communicate its state to ensure sufficient estimation fidelity at the base station while conserving resources?}

To pose this question mathematically, let us denote the estimated state at the remote estimator at time $t$ by $\hat{x}_t \in \mathbb{R}^n$, and the scheduling decision by $\delta_t \in \{0,1\}$. Here, $\delta_t = 1$ indicates a communication by the scheduler. Further, let us define the set of (possibly random) scheduling instants up to time $t$ as
\begin{align}\label{eq:transmission_set}
    {\tt T}_t := \{t_{\ell} \mid \ell = 1, \ldots, n_t\},
\end{align}
where $n_t$ is the total number of communications up to time $t$. 
We assume that the set ${\tt T}_t$ is ordered, i.e., $t_1 < t_2<\cdots<t_{n_t} \le t$.
We have that $\hat x_m = x_m$, if and only if, $m \in {\tt T}_t$. Henceforth, we will let ${\tt T} := \lim_{t\to \infty} {\tt T}_t$. Subsequently, we can define the information available to the scheduler and the estimator at time $t$, respectively, as:
\begin{align}\label{eq:info_sets}
    \mathcal{I}^s_t &:= \{x_k, \delta_k, \hat{x}_k, x_t \mid k \in [t-1]\}, \quad \forall t \geq 1, \nonumber \\
    \mathcal{I}^e_t &:= \{x_m, \delta_k, \hat{x}_{k} \mid m \in {\tt T}_t, k \in [t-1]\}, ~ \forall t \geq 1,
\end{align}
with $\mathcal{I}^s_0 := \{x_0,\delta_0\}$ and $\mathcal{I}^e_0:= \{\delta_0\}$.
Next, we define $\pi$ as the set of measurable scheduling policies according to which decisions $\delta_t$ are made:
\begin{align*}
    \pi = \left\{ \mathbf{P}(\delta_t \mid \mathcal{I}^s_t) \right\}_{\forall t},
\end{align*}
and the set of measurable estimation policies $\mu$, such that:
\begin{align*}
    \mu = \{\mathbf{P}(\hat x_t \mid \mathcal{I}^e_t)\}_{\forall t},
\end{align*}
where $\mathbf{P}(\cdot)$ denotes a Borel-measurable stochastic kernel defined over suitable measurable spaces. The aim is to minimize the following multi-objective cost, expressed as:
\begin{align}\label{eq:cost}
    J(\pi, \mu) := \mathbb{E} \left[ \sum_{t = 0}^{\infty} \gamma^t(\|x_t - \hat{x}_t\|^2_\Gamma + \lambda \delta_t) \right],
\end{align}
by co-designing the scheduler-estimator pair $(\delta,\mu)$. In the above, $\Gamma \succeq 0$ denotes a weighting matrix, $\gamma \in (0,1)$ is the discount factor, and $\lambda > 0$ is a scalar that penalizes communication. Thus, we can formally state the problem as follows\footnote{Instead of the uncontrolled system considered in \eqref{eq:nonlinsystem}, one could indeed consider a controlled dynamical system given by $x_{t+1} = g(x_t, u_t, w_t)$, and subsequently substitute for a stabilizing feedback control $u_t = h(x_t)$ (under suitable conditions of well-posedness to ensure its existence) to arrive at the closed-loop uncontrolled system $x_{t+1} = g(x_t, h(x_t), w_t)=: f(x_t, w_t)$.}.

\begin{problem}\label{problem1}
    We wish to solve the following:
    \begin{align}
        & \inf_{\pi, \mu} J(\pi, \mu) \quad \text{ subject to}\quad \eqref{eq:nonlinsystem}, \eqref{eq:info_sets}.
    \end{align}
\end{problem}

\section{Scheduler-Estimator Characterization}

We begin by stating the following standard result that characterizes the structure of the optimal estimator under a fixed scheduling policy.
\begin{proposition}\cite{molin2017event}\label{thm:opt_estimation}
    Suppose that a scheduling policy $\pi$ is fixed. Then, the optimal estimator for Problem \ref{problem1} is given by the conditional expectation of the state, conditioned on the estimator's information set:
    \begin{align}\label{eq:opt_est}
        \hat{x}_t  = \mu^*(\mathcal{I}^e_t) = \mathbb{E}[x_t \mid \mathcal{I}^e_t], \quad \forall t \geq 0.
    \end{align}
\end{proposition}

Next, to motivate the technical challenges in the design of \eqref{eq:opt_est}, and the optimal scheduling policy $\pi^*$, let us consider a simpler exposition of linear systems in the next subsection.

\subsection{Key Technical Challenges: A Linear System Case Study}
Let us begin by considering a linear-time-invariant system, which is a special case of the nonlinear system \eqref{eq:nonlinsystem}, given as:
\begin{align}\label{eq:lin_system}
    x_{t+1} = Ax_t + w_t, \quad \forall t \geq 0,
\end{align}
where $A \in \mathbb R^{n \times n}$ denotes the system matrix.
Subsequently, by employing the optimal estimator presented in Proposition \ref{thm:opt_estimation} and taking the conditional expectation in \eqref{eq:lin_system}, we obtain the following estimator dynamics:

\begin{align}\label{eq:estimation}
    \hat{x}_{t+1} \!=\! 
    \begin{cases}
        A \mathbb{E}[x_t\mid \mathcal{I}^e_{t+1}] +  \mathbb{E}[w_t \mid \mathcal{I}^e_{t+1}], & \text{if } \delta_{t+1} = 0, \\
        x_{t+1}, & \text{if } \delta_{t+1} = 1,
    \end{cases}
\end{align}
for all $t \geq 0$, with $\hat{x}_0 = (1-\delta_0)\mathbb{E}[x_0] + \delta_0 x_0$.
Note that, when $\delta_{t+1} = 0$, from \eqref{eq:info_sets}, we have $\mathcal{I}^e_{t+1} = \mathcal{I}^e_{t} \cup \{\delta_{t+1}=0\}$. 

Define the estimation error as $e_t := x_t - \hat{x}_t$. Then, using \eqref{eq:lin_system} and \eqref{eq:estimation}, the evolution of the estimation error is given by:
\begin{align}\label{eq:estimation_error}
    e_{t+1} & = x_{t+1} - \hat x_{t+1} =
    \begin{cases}
        A e_t + \hat{w}_t =: e^o_{t+1}, & \text{if } \delta_{t+1} = 0, \\
        0, & \text{if } \delta_{t+1} = 1,
    \end{cases}
\end{align}
with $e_0 = x_0 - \hat x_0$, $\hat{w}_t := w_t - \mathbb{E}[w_t \mid \mathcal{I}^e_{t+1}] + A (\hat x_t - \mathbb E[x_t \mid \mathcal{I}_{t+1}])$, $e^o_0 = x_0 - \mathbb E[x_0]$, and we define $e^o_t$ to be the one-step lookahead error at time $t$, assuming no scheduling (i.e., $\delta_t = 0$).
Accordingly, the cost function in \eqref{eq:cost} can be equivalently expressed as:
\begin{align}\label{eq:est_cost}
    J(\pi, \mu^*) := \mathbb{E} \left[ \sum_{t = 0}^{\infty} \gamma^t ((1-\delta_t)\|e^o_t\|^2_\Gamma + \lambda \delta_t) \right],
\end{align}
We can now restate our objective toward solving Problem \ref{problem1} (for this linear system) as minimizing \eqref{eq:est_cost} over $\pi$ subject to \eqref{eq:estimation}, \eqref{eq:estimation_error} and \eqref{eq:info_sets}. There are, however, two technical challenges associated with the design of a solution to this problem, namely, the computation of the optimal estimator, and the subsequent computation of the scheduling policy.

\subsubsection{Technical Challenge \#1:}
The presence of the conditional expectation terms in \eqref{eq:estimation} renders the estimation dynamics nonlinear, and hence, not expressible in closed form. The challenge becomes even more pronounced for the case of general nonlinear systems. In some cases, however, as presented in the following remarks, this issue can be circumvented but only by imposing additional restrictions over the problem structure, even for the simple case of linear systems.

\begin{remark}[Restrictions over the information sets]\label{rem:separation}
Existing works (see the dissertation by Molin \cite{molin2014optimal} and the references therein and \cite{aggarwal2025interq}) simplify the estimator design by not including the scheduling signal $\delta_t$ in the estimator's information set $\mathcal{I}^e_t$. Thus, one disregards the no-communication events (denoted by $\delta_t = 0$), and uses the approximation:
\begin{align*}
    \mathbb{E}[x_t\mid \mathcal{I}^e_{t+1}] &= \mathbb{E}[x_t\mid \mathcal{I}^e_{t} \cup \{\delta_{t+1}=0\}] \overset{(\dagger)}{\approx} \mathbb{E}[x_t\mid \mathcal{I}^e_{t}] = \hat{x}_t,
\end{align*}
where $(\dagger)$ is precisely where the no-communication events are dropped from the conditional expectation. Additionally, the assumption that $w_t$ is independent across time, and has zero mean, yields $\mathbb{E}[w_t \mid \mathcal{I}^e_{t+1}] = 0$. Consequently, the estimator dynamics in \eqref{eq:estimation} reduce to a piecewise linear form:
    \begin{align*}
        \hat{x}_{t+1} = \delta_{t+1} x_{t+1} + (1 - \delta_{t+1}) A \hat{x}_t,
    \end{align*}
    which can now be computed separately, and recursively for a given scheduling policy. 
\end{remark}

\begin{remark}[Restrictions on noise distribution and policy space]
    Another common restriction in the literature pertains to that on the noise distribution and the scheduler policy space \cite{ramesh2013design,hertneck2025current}. Let us denote the domain of the state space where the scheduler decides not to schedule by $\mathcal{D}$. If $\mathcal{D}$ is symmetric about the origin—e.g., $\mathcal{D} := \{z \in \mathbb{R}^n \mid \|z\| \leq a \text{ for some } a > 0\}$—and the noise distribution is also symmetric over $\mathcal{D}$, the estimator in \eqref{eq:estimation} can again be computed using recursive form as presented in Remark \ref{rem:separation} \cite{maity2020minimal}.
\end{remark}

\begin{remark}[Restriction to scalar linear systems]
    We finally remark that linear piecewise estimation (as in Remark \ref{rem:separation}) is known to be optimal for scalar systems under zero-mean Gaussian noise \cite{imer2010optimal, lipsa2011remote}. In such cases, the absence of communication does not provide additional information for the estimator, and the scheduling policy turns out to be of threshold type, i.e. $\delta = 1$ whenever $|e| \geq \tau$, for a positive scalar $\tau$. However, for multivariate systems, no general result exists, except in certain special cases such as the `innovation signal'-based scheduling policy \cite{maity2020minimal} or the one-step-delay information transmission pattern \cite{soleymani2022value}.
\end{remark}
The above discussion (even though involving linear systems) motivates the first technical challenge in estimation design. In this work, we consider the case of general nonlinear systems (and this appears to be the first such study addressing joint scheduler-estimator design in this context) and do not impose any of the simplifying restrictions noted above. In particular, we retain full information in the estimator, allow for general noise distributions, and do not restrict the policy class, for which we will later resort to a learning-based design after presenting the second technical challenge below.

\subsubsection{Technical Challenge \#2:} Now, we elaborate on the second challenge, which is that of optimal scheduling policy computation. To this end, let us define an MDP ${\tt M} := ({\tt S}, {\tt A}, {\tt P}, {\tt C})$, where ${\tt S}$ is the state space, ${\tt A}$ is the action space, ${\tt P}$ describes the transition dynamics, and ${\tt C}$ defines the per-step cost. In our setting, the MDP state is defined as the lookahead estimation error $e^o_t$, with its dynamics as in~\eqref{eq:estimation_error}, and hence ${\tt S} = \mathbb{R}^n$. The action space ${\tt A} = \{0,1\}$ corresponds to the binary scheduling decisions with the running cost:
\[
{\tt C}(e, \delta) = (1-\delta)\|e^o\|^2_\Gamma + \lambda \delta.
\]

Next, let us fix a scheduling policy $\pi$, and define the state-action value function $Q^{\pi}(e^o,\delta): {\tt S} \times {\tt A} \rightarrow \mathbb R$ of the MDP:
\begin{align*}
    Q^{\pi}(e^o,\delta) \!=\! \mathbb E \left[\sum_{t = 0}^\infty \gamma^t {\tt C}(e^o_t,\delta_t) \mid \pi, e^o_0 = e, \delta_0 = \delta \right].
\end{align*}

 Using the Bellman equation, for any time $t$, we write the optimal state-action value function as
 \begin{align} \label{eq:Q-func}
     Q(e^o_t,\delta_t) & = \inf_{\pi} Q^{\pi}(e^o_t,\delta_t) \nonumber \\
     & = \mathbb E [ {\tt C}(e^o_t, \delta_t) + \gamma \min_{\delta' \in {\tt A}} Q(e^o_{t+1}, \delta')].
 \end{align}
 Subsequently, one may compute the optimal scheduling policy as
 \begin{align}\label{eq:Q_policy}
     \pi^*(\mathcal{I}^s_t) = \arg\min_{\delta_t \in {\tt A}} Q(e^o_t,\delta_t). 
 \end{align}
The challenge, however, is that in the given definition, the closed-form expression of the state-action value function remains unknown, making it impossible to compute the scheduling policy exactly.

\begin{remark}
    We remark that there have been multiple attempts to compute the optimal scheduling policy, for the simpler case of linear systems \cite{soleymani2021value,maity2020minimal,soleymani2022value,hertneck2025current}. For instance, the work \cite{soleymani2022value} computes an optimal scheduling policy for a (one-step delayed) linear system only to discover that the policy depends on the value function of the system, and hence, cannot be implemented without appropriate approximations. To alleviate this issue, a recent work \cite{aggarwal2025interq} attempts to compute the scheduling policy using a deep Q-learning algorithm, albeit by approximating the estimator to its piecewise linear form as discussed earlier.
\end{remark}
Building on the preceding discussions on the key challenges, this work is the first to consider the co-design problem for nonlinear systems to explicitly show that communication gaps (i.e., no-scheduling events) can convey implicit information that improves estimation performance in multivariate systems, and without imposing any restrictions as presented before. However, due to the underlying intractability within the estimator, and the coupled scheduling policy design, we resort to a DRL-based co-design framework within an actor-critic architecture: the actor (scheduler deep neural network (DNN)) selects a communication policy, while the critic (estimator DNN) evaluates and updates the state estimate accordingly. We detail this approach in the following section.

\section{Method \& Algorithm}
We now describe the alternating training algorithm used to learn both the communication scheduling policy and the state estimator dynamics. The scheduler is modeled using a neural network function approximator which is then trained via PPO with the clipped surrogate objective \cite{schulman2017proximal}.
The estimator is also modeled using a neural network function approximator, which is trained to predict the system state under sporadic observations. By alternating between training the scheduler (while holding the estimator fixed) and training the estimator (with the scheduler fixed), we enable coordinated learning to balance estimation accuracy and communication cost.

\subsubsection{PPO subroutine.} We briefly outline the PPO algorithm \cite{schulman2017proximal}, which serves as a subroutine in our framework. PPO is a first-order reinforcement learning method that addresses instability in traditional policy gradient approaches by limiting large updates via a clipped surrogate objective, thereby offering a simpler and more efficient alternative to TRPO \cite{schulman2015trust}.

The foundations of PPO lie within the standard technique of estimating the gradient of expected returns in policy gradient formulation:
\begin{equation*}
    \hat{g} = \mathbb{E}_t \left[ \nabla_\theta \log \pi_\theta(\delta_t \mid e^o_t) \hat{A}^v_t \right],
\end{equation*}
which corresponds to maximizing the objective function:
\begin{equation*}
    L^{\rm PG}(\theta) = \mathbb{E}_t \left[ \log \pi_\theta(\delta_t \mid e^o_t) \hat{A}^v_t \right].
\end{equation*}
Here, $\pi_\theta$ is the current policy parameterized by $\theta$, $\hat{A}^v_t$ is an estimator of the advantage function \cite{schulman2017proximal}, and $\mathbb E_t$ denotes the empirical average taken over a finite number of samples. 
To ensure stability, PPO introduces a clipped objective:
\begin{equation*}
    L^{\text{CLIP}}(\theta) = \mathbb{E}_t \left[ \min \left( z_t(\theta) \hat{A}^v_t,~ \text{clip}(z_t(\theta), 1 \pm \epsilon) \hat{A}^v_t \right) \right],
\end{equation*}
where $z_t(\theta) = \frac{\pi_\theta(\delta_t \mid e^o_t)}{\pi_{\theta_{\text{old}}}(\delta_t \mid e^o_t)}$ is the policy ratio between the current and the old policy, and $1 \gg\epsilon>0$ is a hyperparameter. This clipping limits the incentive for overly large updates to remain within a `trust' region, promoting more stable learning. The resulting algorithm alternates between sampling data using the current policy and optimizing the clipped objective over multiple epochs of stochastic gradient ascent.

\subsection{Proposed Algorithm: \texttt{CALM}}
In light of the above discussion on PPO, we propose \texttt{CALM}: \underline{C}ommunication-\underline{A}ware \underline{L}earning and \underline{M}onitoring (presented in Algorithm~\ref{Alg:CALM}) to jointly construct the scheduler and the estimator. The algorithm alternates between updating the communication scheduling policy via PPO, and subsequently refining the estimator using stochastic gradient descent (SGD). The procedure begins by initializing three neural networks: the estimator network, the policy network, and the value function network. An outer loop (line~1) is then executed, within which the scheduler is trained (lines~2--15) followed by the estimator training (lines~16--26). To enable the estimator to more effectively capture temporal dynamics, we incorporate an \emph{age-of-information} (AoI) feature as an additional input during training. Specifically, the AoI is defined as the time elapsed since the last transmission from the scheduler to the estimator. If the current time is $t$, and the most recent communication occurred at time $t_{n_t}$, then $\text{AoI} = t - t_{n_t}$, where $n_t$ is defined as in \eqref{eq:transmission_set}.
The training is performed in a centralized manner, i.e., each DM maintains a local copy of the trained network of the other DM. During execution, however, both DMs act solely based on the information available to them, as defined in \ref{eq:info_sets}. Further, to ensure faster learning, we choose 
\begin{align*}
    \mu^*(\mathcal{I}^e_{t+1}) = \begin{cases}
    f(\hat{x}_t,0) + \xi_\psi (\hat{x}_t, t-t_{n_t}), ~ & \delta_{t+1} = 0, \\
    x_{t+1}, & \delta_{t+1} = 1,
\end{cases}
\end{align*}
i.e., the estimator DNN (denoted by $\xi_\psi(\cdot)$) only estimates the residual sum: $\mathbb{E}[f(x_{t},w_t) \mid \mathcal{I}^e_{t+1}] - f(\hat{x}_t,0)$.

\begin{algorithm2e}[h!]
\caption{{\texttt{CALM}: } PPO-driven Alternating Scheduler-Estimator Training Algorithm}\label{Alg:CALM}
\KwIn{Initial estimator $\xi_{\psi_0}$, scheduling policy $\pi_{\theta_0}$, value function $V_{\phi_0}$, noise distributions $\mathcal{P}, \mathcal{P}_0$, cost weight $\lambda$, rollout length $T$}
\KwOut{Trained policy $\pi_\theta$ and estimator $\xi_\psi$}

\For{each outer iteration $i = 1$ to $N$}{
    \tcp{Train Scheduler using PPO with fixed Estimator}
    \For{each PPO epoch}{
        \For{each trajectory rollout}{
            Initialize $x_0 \sim$ Uniform[-1,1], $\hat{x}_0 \gets \mathbb E [x_0]$, $e_0 = x_0 - \hat{x}_0$, $t_{n_t} \gets 0$\;
            \For{$t = 0$ to $T-1$}{
                Observe error $e_t$\;
                Sample action $\delta_t \sim \pi_\theta(e_t)$\;
                \If{$\delta_t = 1$}{
                    $\hat{x}_t \gets x_t$\;
                    $t_{n_t} \gets t$\;
                }
                
                Compute reward $r_t = -\left(\|x_t - \hat{x}_t\|_\Gamma^2 + \lambda \cdot \mathbb{I}_{\{\delta_t = 1\}}\right)$\;

                
                $x_{t+1} \gets f(x_t, w_t)$\;
                

                $\hat{x}_{t+1} \gets f(\hat{x}_t, 0) + \xi_{\psi}(\hat{x}_t, t - t_{n_t})$\;
                
            }
            Store $(e_t, \delta_t, \log\pi_\theta(\delta_t|e_t), z_t, V_\phi(e_t))$ and compute advantage estimate $\hat A^v_t$\;
        }
        Perform PPO update on $\theta$ and $\phi$ using collected trajectories
    }

    \tcp{Train Estimator using Fixed Policy}
    \For{each estimator training epoch}{
        Initialize $x_0 \sim$ Unif[-1,1], $\hat{x}_0 \gets \mathbb E[x_0]$, $t_{n_t} \gets 0$\;
        \For{$t = 0$ to $T-1$}{
            Observe error $e_t = x_t - \hat{x}_t$\;
            Sample action $\delta_t \sim \pi_\theta(e_t)$\;
            \If{$\delta_t = 1$}{
                $\hat{x}_t \gets x_t$, $t_{n_t} \gets t$\;
            }
            
            Accumulate loss ${L}$: ${L} = \gamma  L \mathrel{+} \|x_t - \hat{x}_t\|_\Gamma^2 + \lambda \cdot \mathbb{I}_{\{\delta_t = 1\}}$ 

            $x_{t+1} \gets f(x_t, w_t)$\;

            {
                $\hat{x}_{t+1} \gets f(\hat{x}_t, 0) + \xi_{\psi}(\hat{x}_t, t - t_{n_t})$\;
            }
        }
        Update estimator parameters $\psi$ using $\nabla_\psi {L}$\;
    }
}
\end{algorithm2e}

\section{Experiments and Analysis}

\subsubsection{Benchmark Details:} 
We evaluate the proposed \texttt{CALM} framework across three standard control tasks: \textit{inverted pendulum stabilization, Van der Pol oscillator},  \textit{trajectory tracking} problem, and a Boeing flight control system. For all experiments, we use the p-mode Gaussian mixture to model the noise distribution, which naturally occurs in diverse mechanical systems such as flight control, human-robot synergies, etc. \cite{zhang2023extreme,hernandez2024bayesian,iannamorelli2025adaptive}. Additional implementation and environment-specific details, and signal trajectories corresponding to all experiments are provided in the Appendix.

\subsection{Results}

\subsubsection{Inverted Pendulum.}
\begin{figure}
    \centering
    \begin{minipage}{0.49\columnwidth}
        \centering
        \includegraphics[width=0.8\textwidth]{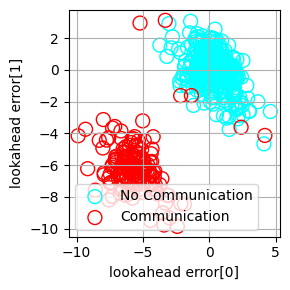} 
    \end{minipage}\hfill
    \begin{minipage}{0.49\columnwidth}
        \centering
        \includegraphics[width=0.8\textwidth]{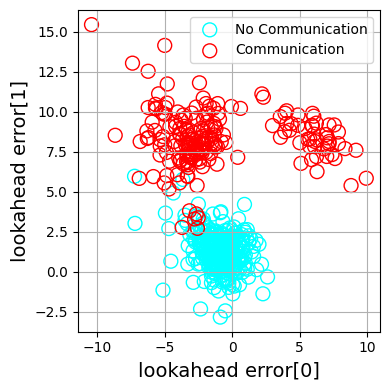} 
    \end{minipage}
    \caption{\small{Scheduling landscapes for a 2-mode GMM (left) and 3-mode GMM (right) for the inverted pendulum system (red denotes transmissions while cyan denotes silence).}}
    	\label{Fig:scatter_IP}
 	\vspace{-0.2cm}
\end{figure}

For our first experiment, we consider an inverted pendulum system subject to noise drawn from a two-mode Gaussian mixture model (GMM), as shown in (the left subfigure of) Fig.~\ref{Fig:scatter_IP}. The GMM has component means at $(-3,-3)$ and $(3,3)$, equal covariance matrices ${\diag}(0.5, 0.5)$, and respective mixture weights of $0.3$ and $0.7$. 
For this example, we also perform an ablation study, where we evaluate performance under a three-mode GMM with component means $(-3,-3)$, $(-5,4)$, and $(4,4)$, and covariance matrices $\mathrm{diag}(0.5, 0.5)$, $\begin{pmatrix} 1.0 & 0.8 \\ 0.8 & 1.0 \end{pmatrix}$, and $\begin{pmatrix} ~~0.6 & \!\!\!-0.3 \\ -0.3 & ~0.5 \end{pmatrix}$, with associated mixture weights $(0.6, 0.3, 0.1)$. The communication cost is set to $\lambda = 45$, with further experimental details provided in the Appendix. In Fig.~\ref{Fig:scatter_IP}, we visualize the scheduling policy learned by \texttt{CALM} as a function of the lookahead estimation error.
The resulting decision landscape clearly partitions the state space into two regions: one favoring communication (in red), and the other favoring silence (in cyan).

\textbf{Silence that speaks:} This behavior reflects a form of implicit coordination between the scheduler and the estimator. When the error in Fig.~\ref{Fig:scatter_IP} (on the left) is likely drawn from the positively shifted GMM mode, the scheduler chooses not to transmit (cyan), and the estimator correctly infers this and incorporates the corresponding mode mean into its estimate update. Conversely, when the samples are likely drawn from the negatively shifted mode, the scheduler actively communicates (red), and the estimator uses the actual state. This is a clear instance of ``communication via silence'' (or a signaling type behavior), where the absence of a transmission conveys informative structure about the underlying noise distribution. Similarly, in the right subfigure in Fig.~\ref{Fig:scatter_IP}, when samples are drawn from the component in the third quadrant, no communication takes place while it does when sampled from the other two regions.

In contrast, a baseline linear estimator (i.e., the best Kalman filter)---ignorant of this implicit information—simply adds the overall GMM mean without any knowledge of the noise structure, leading to higher overall incurred cost. Specifically, corresponding to Fig.~\ref{Fig:scatter_IP} (left), \texttt{CALM} resulted in 168 communications (for $T=500$) with a total cost of 8355.16 as per \eqref{eq:est_cost}, while the best linear baseline registered 256 transmissions and incurred a cost of 13386.55. Similarly, for Fig.~\ref{Fig:scatter_IP} (right), \texttt{CALM} resulted in 199 communications with a total cost of 10385.45, while the linear baseline registered 174 transmissions and incurred a cost of 11174.44. 
Results corresponding to the baseline are presented in the Appendix.
These results underscore the central premise of our work: silence can carry structured information, which, if leveraged properly, can significantly improve estimation performance.

\subsubsection{Van der Pol (VdP) oscillator.}
\begin{figure}
    \centering
    \begin{minipage}{0.49\columnwidth}
        \centering
        \includegraphics[width=0.8\textwidth]{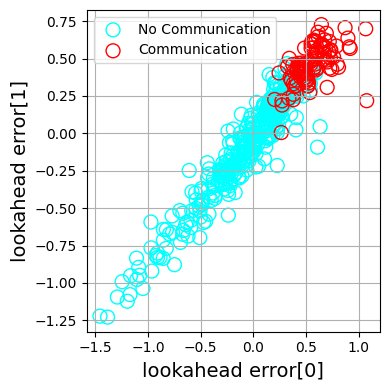} 
    \end{minipage}\hfill
    \begin{minipage}{0.49\columnwidth}
        \centering
        \includegraphics[width=0.8\textwidth]{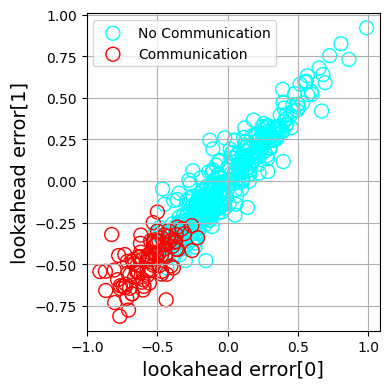} 
    \end{minipage}
    \caption{\small{Scheduling landscapes for different weights of the 2-modes of GMM for VdP system: weight vector = (0.3, 0.7) on the left and weight vector = (0.7, 0.3) on the right.}}
    	\label{Fig:scatter_vdp}
 	\vspace{-0.2cm}
\end{figure}

Our next case study involves the VdP oscillator, a nonlinear dynamical system of dimension 2. As in the inverted pendulum case, we consider a two-mode GMM with component means at $(-5,-4)$ and $(4,5)$, both having identical covariance matrices ${\diag}(0.5,0.5)$. 
In Fig.~\ref{Fig:scatter_vdp}, we visualize the scheduling landscape generated by \texttt{CALM} as a function of the lookahead estimation error.

Similar to the inverted pendulum case, we again observe the emergence of an agreement between the scheduler and the estimator, resulting in a clear partitioning of the decision space. Further, the positively shifted GMM mode is selected for non-communication in the left subfigure of Fig.~\ref{Fig:scatter_vdp} due to its higher likelihood (weight of 0.7), thereby minimizing the need for explicit communication while preserving estimation fidelity. As an ablation study, when the GMM weights are reversed, the scheduler-estimator coordination also inverts. This results in a flipped decision landscape, where the previously silent mode now triggers communication and vice versa, as illustrated in  the right subfigure of Fig.~\ref{Fig:scatter_vdp}.

\subsubsection{Robot trajectory tracking.}
\begin{figure}[t]
    \centering
    \begin{minipage}{0.49\columnwidth}
        \centering
        \includegraphics[width=0.8\textwidth]{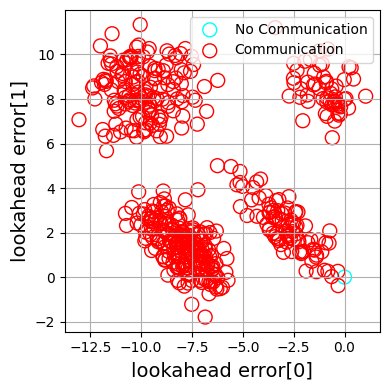} 
    \end{minipage}\hfill
    \begin{minipage}{0.49\columnwidth}
        \centering
        \includegraphics[width=0.8\textwidth]{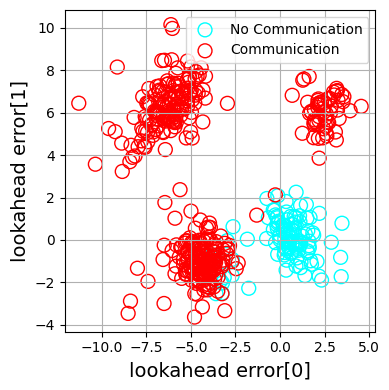} 
    \end{minipage}
    \begin{minipage}{0.49\columnwidth}
        \centering
        \includegraphics[width=0.8\textwidth]{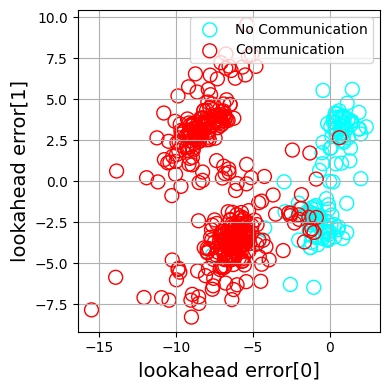} 
    \end{minipage}\hfill
    \begin{minipage}{0.49\columnwidth}
        \centering
        \includegraphics[width=0.8\textwidth]{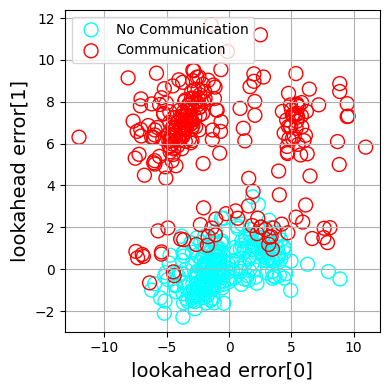} 
    \end{minipage}
    \caption{\small{Variation of scheduling landscape with communication cost $\lambda$ for the trajectory tracking experiment: $\lambda=15$ (top left), $\lambda=30$ (top right), $\lambda=40$ (bottom-left) and $\lambda=70$ (bottom right).}}
    	\label{Fig:scatter_pt}
 	\vspace{-0.3cm}
\end{figure}

Our next experiment involves a trajectory tracking system in 2 dimensions, similar to the aforementioned Rover example. For this study, we fix a 4-mode GMM with mean vectors $(-3,-3), (-5,4),$ $ (2,-2), (4,4)$ and covariance matrices {\small  $\diag(0.5,0.5)$, $\begin{pmatrix}
    1 & 0.8 \\
    0.8 & 1
\end{pmatrix}$, $\begin{pmatrix}
    ~0.6 & \!\!-0.3 \\
    \!\!-0.3 & 0.5
\end{pmatrix}$, and $\begin{pmatrix}
    0.3 & 0.1 \\
    0.1 & 0.4
\end{pmatrix}$}.
The weight vector was set to $(0.4, 0.3, 0.2, 0.1)$. In Fig. \ref{Fig:scatter_pt}, we investigate how the learned scheduling landscape evolves as a function of the communication cost parameter $\lambda$. For the smallest cost setting, $\lambda = 15$ (top-left), the system has sufficient communication resources to enable communication (indicated by red) for all the four modes.
As the cost increases to $\lambda = 30$ (top-right), we observe a no-communication region (indicated in cyan) in the fourth quadrant while communicating regions in the other three quadrants.

At a more restrictive communication cost of $\lambda = 40$ (bottom-left), the scheduler now begins to communicate in only two of the four GMM modes, with increased selectivity under tighter communication constraints. Finally, when the communication cost becomes very high ($\lambda = 70$ (bottom-right)), the budget becomes too limited, and hence, we see a denser cyan region of no-communications compared to the lighter red regions (indicating communication). Further, the communication region also swaps to the first quadrant (from the third quadrant for the case with $\lambda=40$) since it has a lesser likelihood of being sampled (only 0.1 compared to 0.4 in case of the mode in the third quadrant).
This adaptive trade-off
highlights the algorithm’s ability to allocate communication
resources intelligently based on both statistical structure
(e.g., covariance and probability mass) and cost parameter.
This trend can be further verified by the tuple of overall
cost and number of transmissions, which was observed
to be $(7301.67, 499), (11683.64, 386), (16160.71, 374)$ and$(19002.01, 249)$ for increasing value of $\lambda$ in  Fig. \ref{Fig:scatter_pt} over a horizon of 500 units. The corresponding signal trajectories for each case are provided in the Appendix.

\subsubsection{Boeing flight control} Our final experimental setup is that of a 4-dimensional Boeing flight control system \cite{boyd2018introduction,ataei2025qsid}. We take 2-mode GMM with mean values $\begin{pmatrix}
    -5.0 & -4.0 & -3.0 & -2.0 \\
    4 & 5 & 3.0 & 2.0
\end{pmatrix}^\top$, covariance matrix 

$\begin{pmatrix}
    \mathrm{diag} (0.1 & 0.2 & 0.3 &, 0.4) \\
    \mathrm{diag} (0.4 & 0.1 & 0.3 & 0.2)
\end{pmatrix}$, and the weight vector as $(0.3, 0.7)$. The scatter plot demonstrating the scheduling landscape (for $\lambda=45$) and the implicit scheduler-estimator agreement for the GMM modes, is presented in Fig. \ref{Fig:scatter_bo} along with its corresponding signal trajectories in Fig. \ref{Fig:traj_bo} in the Appendix.

\begin{figure}
	\centering	\includegraphics[width=0.4\columnwidth ]{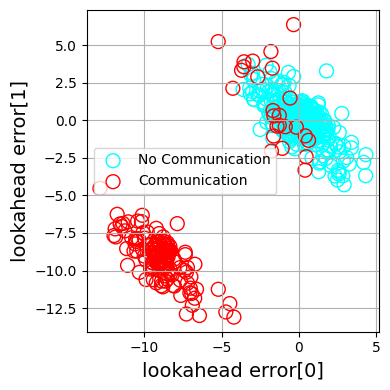}
	\caption{\small{Scheduling landscape for the Boeing flight control system showing clear partitioning between communication and no-communication events.
	}}
	\label{Fig:scatter_bo}
 	\vspace{-0.2cm}
\end{figure}

\subsection{Baseline Comparison} 
\begin{figure}[h!]
    \centering
    \begin{minipage}{0.5\columnwidth}
        \centering
        \includegraphics[width=0.8\textwidth]{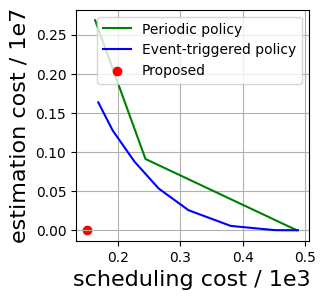} 
            \label{Fig:a}
    \end{minipage}\hfill
    \begin{minipage}{0.5\columnwidth}
        \centering
        \includegraphics[width=0.8\textwidth]{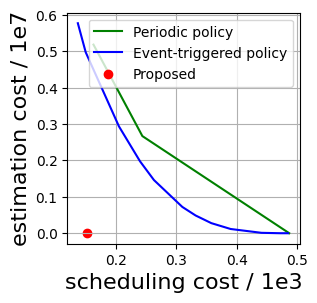}
        \label{Fig:b}
    \end{minipage}
    \vspace{-.25cm}
    \caption{\small{Estimation-scheduling trade-off for a fixed estimator for tracking (left) and inverted pendulum problem (right).}}
    	\label{Fig:tradeoff1}
 	\vspace{-0.3cm}
\end{figure}

In this study, we fix the estimator to that trained using \texttt{CALM}, and compare the Pareto-fronts of the estimation-scheduling trade-offs produced using the following baselines (see Fig.~\ref{Fig:tradeoff1}).

\begin{itemize}
    \item \textbf{Periodic Scheduling:} A fixed-period transmission scheme in which the plant communicates with the estimator at regular integer time intervals (in an open-loop manner). We evaluate period lengths of 1, 2 and 3, to observe trade-offs between communication frequency and estimation performance.
    
    \item \textbf{Event-Triggered Policy:} A threshold-based feedback scheduling policy of the form $\|e_t\|^2 \geq \tau$, where $\tau > 0$ is a design parameter. Communication occurs only when the estimation error exceeds the threshold.
    
\end{itemize}

From Fig.~\ref{Fig:tradeoff1} (for $\lambda=45$ on the tracking and pendulum problem), we observe that the trade-off performance of the proposed algorithm \texttt{CALM} (over two different sets of noise realizations) is better than the baselines on the periodic policy, and the event-triggered threshold policy. 

\section{Conclusion}

We proposed a novel algorithmic framework for the joint learning of communication schedulers and state estimators in stochastic dynamical systems, using neural networks as function approximators. The approach, applicable to general nonlinear systems, alternates between training the estimator and the scheduler, the latter of which is optimized using proximal policy optimization. Our results highlight a key insight: communication silence---i.e., no-communication events---implicitly carries information that can be leveraged to enhance estimation accuracy, and close a long-standing gap in co-design methodologies. Extensive experiments across standard benchmarks, including inverted pendulum stabilization, Van der Pol oscillator, and robot trajectory tracking systems, reveal that our method consistently outperforms classical estimators and widely used heuristic scheduling strategies such as periodic and event-triggered ones.

\bibliography{refs}

@article{marschak1955elements,
  title={Elements for a theory of teams},
  author={Marschak, Jakob},
  journal={Management Science},
  volume={1},
  number={2},
  pages={127--137},
  year={1955},
  publisher={INFORMS}
}

@article{imer2010optimal,
  title={Optimal estimation with limited measurements},
  author={Imer, Orhan C and Ba{\c s}ar, Tamer},
  journal={International Journal of Systems, Control and Communications},
  volume={2},
  number={1-3},
  pages={5--29},
  year={2010},
  publisher={Inderscience Publishers}
}

@article{soleymani2022value,
  title={Value of information in feedback control: Global optimality},
  author={Soleymani, Touraj and Baras, John S and Hirche, Sandra and Johansson, Karl H},
  journal={IEEE Transactions on Automatic Control},
  volume={68},
  number={6},
  pages={3641--3647},
  year={2022},
  publisher={IEEE}
}

@article{lipsa2011remote,
  title={Remote state estimation with communication costs for first-order {LTI} systems},
  author={Lipsa, Gabriel M and Martins, Nuno C},
  journal={IEEE Transactions on Automatic Control},
  volume={56},
  number={9},
  pages={2013--2025},
  year={2011},
  publisher={IEEE}
}

@article{chang2019neural,
  title={Neural {L}yapunov control},
  author={Chang, Ya-Chien and Roohi, Nima and Gao, Sicun},
  journal={Advances in Neural Information Processing Systems},
  volume={32},
  year={2019}
}

@article{zhou2022neural,
  title={Neural {L}yapunov control of unknown nonlinear systems with stability guarantees},
  author={Zhou, Ruikun and Quartz, Thanin and De Sterck, Hans and Liu, Jun},
  journal={Advances in Neural Information Processing Systems},
  volume={35},
  pages={29113--29125},
  year={2022}
}

@article{wu2023neural,
  title={Neural {L}yapunov control for discrete-time systems},
  author={Wu, Junlin and Clark, Andrew and Kantaros, Yiannis and Vorobeychik, Yevgeniy},
  journal={Advances in Neural Information Processing Systems},
  volume={36},
  pages={2939--2955},
  year={2023}
}

@article{aggarwal2025interq,
  title={{InterQ}: A {DQN} Framework for Optimal Intermittent Control},
  author={Aggarwal, Shubham and Maity, Dipankar and Ba{\c{s}}ar, Tamer},
  journal={IEEE Control Systems Letters},
  year={2025},
  publisher={IEEE}
}

@book{khalil2002nonlinear,
  title={Nonlinear Systems},
  author={Khalil, Hassan K},
  volume={3},
  year={2002},
  publisher={Prentice Hall, Upper Saddle River, NJ}
}

@article{ataei2025qsid,
  title={QSID-MPC: Model Predictive Control with System Identification from Quantized Data},
  author={Ataei, Shahab and Maity, Dipankar and Goswami, Debdipta},
  journal={arXiv preprint arXiv:2503.19102},
  year={2025}
}

@article{radner1962team,
	title={Team decision problems},
	author={Radner, Roy},
	journal={The Annals of Mathematical Statistics},
	volume={33},
	number={3},
	pages={857--881},
	year={1962},
	publisher={JSTOR}
}

@INBOOK{feldbaum1961dual,
  author={A.A., Feldbaum},
  booktitle={Control Theory: Twenty-Five Seminal Papers (1932-1981)}, 
  title={Dual Control Theory}, 
  year={2001},
  volume={},
  number={},
  pages={181-196},
  doi={10.1109/9780470544334.ch10}}

@book{yuksel2013stochastic,
  title={Stochastic Networked Control Systems: Stabilization and Optimization under Information Constraints},
  author={Y{\"u}ksel, Serdar and Ba{\c{s}}ar, Tamer},
  year={2013},
  publisher={Springer Science \& Business Media}
}

@phdthesis{molin2014optimal,
	title={Optimal Event-Triggered Control with Communication Constraints},
	author={Molin, Adam},
	year={2014},
	school={Technische Universit{\"a}t M{\"u}nchen}
}

@article{maity2020minimal,
  title={Minimal feedback optimal control of linear-quadratic-{G}aussian systems: no communication is also a communication},
  author={Maity, Dipankar and Baras, John S},
  journal={IFAC-PapersOnLine},
  volume={53},
  number={2},
  pages={2201--2207},
  year={2020},
  publisher={Elsevier}
}

@article{thilina2013machine,
  title={Machine learning techniques for cooperative spectrum sensing in cognitive radio networks},
  author={Thilina, Karaputugala Madushan and Choi, Kae Won and Saquib, Nazmus and Hossain, Ekram},
  journal={IEEE Journal on Selected Areas in Communications},
  volume={31},
  number={11},
  pages={2209--2221},
  year={2013},
  publisher={IEEE}
}

@book{soderstrom2012discrete,
  title={Discrete-Time Stochastic Systems: Estimation and Control},
  author={S{\"o}derstr{\"o}m, Torsten},
  year={2012},
  publisher={Springer Science \& Business Media}
}

@article{ramesh2013design,
  title={Design of state-based schedulers for a network of control loops},
  author={Ramesh, Chithrupa and Sandberg, Henrik and Johansson, Karl H},
  journal={IEEE Transactions on Automatic Control},
  volume={58},
  number={8},
  pages={1962--1975},
  year={2013},
  publisher={IEEE}
}

@inproceedings{eisen2022communication,
  title={Communication-control co-design in wireless edge industrial systems},
  author={Eisen, Mark and Shukla, Santosh and Cavalcanti, Dave and Baxi, Amit S},
  booktitle={2022 IEEE 18th International Conference on Factory Communication Systems (WFCS)},
  pages={1--8},
  year={2022},
  organization={IEEE}
}

@article{wang2024infinite,
  title={Infinite-horizon optimal scheduling for feedback control},
  author={Wang, Siyi and Hirche, Sandra},
  journal={arXiv preprint arXiv:2402.08819},
  year={2024}
}

@article{hertneck2025current,
  title={Current trends and future directions in event-based control},
  author={Hertneck, Michael and Meister, David and Allg{\"o}wer, Frank},
  journal={arXiv preprint arXiv:2505.22378},
  year={2025}
}

@article{soleymani2021value,
  title={Value of information in feedback control: Quantification},
  author={Soleymani, Touraj and Baras, John S and Hirche, Sandra},
  journal={IEEE Transactions on Automatic Control},
  volume={67},
  number={7},
  pages={3730--3737},
  year={2021},
  publisher={IEEE}
}

@article{schulman2017proximal,
  title={Proximal policy optimization algorithms},
  author={Schulman, John and Wolski, Filip and Dhariwal, Prafulla and Radford, Alec and Klimov, Oleg},
  journal={arXiv preprint arXiv:1707.06347},
  year={2017}
}

@inproceedings{schulman2015trust,
  title={Trust region policy optimization},
  author={Schulman, John and Levine, Sergey and Abbeel, Pieter and Jordan, Michael and Moritz, Philipp},
  booktitle={International Conference on Machine Learning},
  pages={1889--1897},
  year={2015},
  organization={PMLR}
}

@article{iannamorelli2025adaptive,
  title={Adaptive Gaussian mixture filtering for multi-sensor maneuvering cislunar space object tracking},
  author={Iannamorelli, John L and LeGrand, Keith A},
  journal={The Journal of the Astronautical Sciences},
  volume={72},
  number={1},
  pages={2},
  year={2025},
  publisher={Springer}
}

@article{zhang2023extreme,
  title={Extreme wind turbine response extrapolation with the Gaussian mixture model},
  author={Zhang, Xiaodong and Dimitrov, Nikolay},
  journal={Wind Energy Science},
  volume={8},
  number={10},
  pages={1613--1623},
  year={2023},
  publisher={Copernicus Publications G{\"o}ttingen, Germany}
}

@article{hernandez2024bayesian,
  title={Bayesian intention for enhanced human robot collaboration},
  author={Hernandez-Cruz, Vanessa and Zhang, Xiaotong and Youcef-Toumi, Kamal},
  journal={arXiv preprint arXiv:2410.00302},
  year={2024}
}

@article{kalman1960new,
  title={A new approach to linear filtering and prediction problems},
  author={Kalman, Rudolph Emil},
  year={1960}
}

@article{molin2017event,
  title={Event-triggered state estimation: An iterative algorithm and optimality properties},
  author={Molin, Adam and Hirche, Sandra},
  journal={IEEE Transactions on Automatic Control},
  volume={62},
  number={11},
  pages={5939--5946},
  year={2017},
  publisher={IEEE}
}

@book{boyd2018introduction,
  title={Introduction to Applied Linear Algebra: Vectors, Matrices, and Least Squares},
  author={Boyd, Stephen and Vandenberghe, Lieven},
  year={2018},
  publisher={Cambridge University Press}
}

\section*{Appendix: Supplementary Material}

\section{Further Details about Experiments}

\subsection{System Dynamics}
We describe all the system dynamics in continuous-time. Subsequently, we discretize them by setting the discretization interval to 0.05 seconds in all experiments.

\subsubsection{Inverted Pendulum} We use the following stochastic discrete-time dynamics model for the inverted pendulum system \cite{chang2019neural,zhou2022neural,wu2023neural}:
\begin{align*}
    x_{t+1} = \begin{pmatrix}
        1 & \varepsilon \\
        \frac{g}{\ell}\varepsilon & 1-\frac{b}{m\ell^2}\varepsilon
    \end{pmatrix}x_t + \begin{pmatrix}
        0 \\
        \frac{1}{m \ell^2}\varepsilon
    \end{pmatrix} u_t + w_t,
\end{align*}
where the state $x:=[\theta, \dot \theta]^\top$ captures the angle $\theta$ and the angular velocity $\dot \theta$. The parameter $g$ denotes gravity, $m$ denotes mass of the ball, $b$ denotes friction coefficient, and $\ell$ denotes pendulum length. For all experiments, we take $g = 9.81$, $m = 0.15$, $b = 0.1$, $\ell = 0.5$, and $\varepsilon = 0.05$.

\subsubsection{Van der Pol (VdP) oscillator} We use the following nonlinear dynamics for the VdP oscillator \cite{khalil2002nonlinear}:
\begin{align*}
    (x_1)_{t+1} &= (x_1)_t + \varepsilon[(x_2)_t + (w_1)_t]\\
    (x_2)_{t+1} &= (x_2)_t + \varepsilon[\mu(1-(x_1)_t^2)(x_2)_t - (x_2)_t + (w_2)_t]
\end{align*}
with $x_1$ and $x_2$ denoting the position and velocity of the second-order system with the state $x = [x_1, ~x_2]^\top$. Further, $\mu >0$ denotes the damping strength. For all experiments, we take the parameters as $\mu = 0.025$ and $\varepsilon = 0.05$.

\subsubsection{Robot trajectory tracking system} We use the following discretized model for trajectory-tracking system \cite{wu2023neural}:
\begin{align*}
    x_{t+1} = \begin{pmatrix}
        1 & 2 \\
        0 & 1 - 0.04 \varepsilon
    \end{pmatrix} x_t + \begin{pmatrix}
        0 \\
        \varepsilon
    \end{pmatrix} u_t + w_t,
\end{align*}
where the state $x$ constitutes the distance and the angle errors from the corresponding reference values, the control input constitutes the steering angle $\delta$. For all experiments, we take $\varepsilon = 0.05$.

\subsubsection{Boeing 747 flight control}
We use the following 4-dimensional longitudinal flight control system of the Boeing 747 aircraft in steady level flight at an altitude of 40,000 ft, a speed 774 ft/s, and a time unit of one second \cite{boyd2018introduction,ataei2025qsid}:
\begin{align*}
    x_{t+1} = & \begin{pmatrix}
    0.99 & 0.03 & -0.02 & -0.32 \\
    0.01 & 0.47 & 4.7 & 0.0 \\
    0.02 & -0.06 & 0.40 & 0.0 \\
    0.01 & -0.04 & 0.72 & 0.99
    \end{pmatrix} x_t + \begin{pmatrix}
        0.01 & 0.99 \\
        -3.44 & 1.66 \\
        -0.83 & 0.44 \\
        -0.47 & 0.25
    \end{pmatrix} u_t + w_t
\end{align*}
The state $x_t$ constitutes deviations (from the nominal values) of the velocity along the aircraft body axis, velocity perpendicular to the body axis, angle of the body above horizontal, and derivative of the angle of the body of the body axis. The control input constitutes the deviations of the elevator angle and the engine thrust, from the nominal values.

\subsection{Controlled systems}
For systems with explicit control inputs such as the inverted pendulum, trajectory tracking, and the flight control ones, we apply the infinite-horizon linear-quadratic state feedback control solution. Precisely, for given matrices $Q \succeq 0$ and $R \succ 0$, we apply the control input:
\[
u_t = K x_t
\]
at the dynamical system while at the estimator, which has access to only the best estimate of the state $\hat x_t$ at time $t$, we apply the input:
\[
u_t = K \hat x_t.
\]
Here, we have
\[
K = - \gamma (P + \gamma B^\top R B)^{-1} B^\top P A
\]
and $P \succeq 0$ is the unique positive semi-definite solution to the algebraic Riccati equation:
\[
P = \gamma A^\top P A + Q - \gamma^2 A^\top P B (B^\top R B + P)^{-1} B^\top P A.
\]
For all experiments, we used both Q and R to be identity matrices of suitable dimensions and the discount factor to be 0.9999.

\subsection{Training Hyperparameters}
The training parameters for \texttt{CALM} were set to the following:

Estimator NN activation function: ReLU

PPO network activation function: ReLU

Optimizer for Estimator NN: Adam with weight decay

Optimizer for PPO: Adam

learning rate for both networks: 1e-3

GAE lambda = 0.9


PPO clipping parameter = 0.2

Number of outer epochs: 10

Number of inner epochs for PPO: 80

Number of inner epochs for estimator: 150

Number of epochs for scheduler policy training (PPO) with fixed linear estimator: 1000

Trajectory horizon length: 80

\section{Extended Experiments}
\subsection{Inverted Pendulum}
\begin{figure}[h!]
	\centering	\includegraphics[width=0.9\columnwidth ]{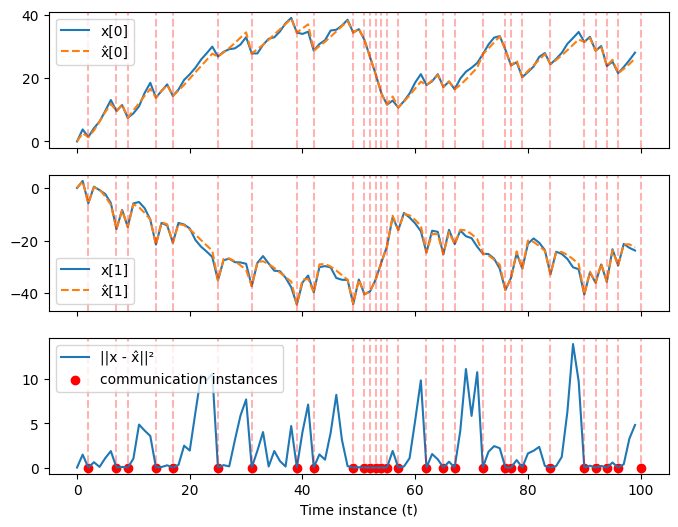}
 \vspace{-0.4cm}
	\caption{\small{Signal trajectories with communication instances (red markers) corresponding to the left subfigure in Fig.~\ref{Fig:scatter_IP} for the inverted pendulum.
	}}
	\label{Fig:traj_IP}
\end{figure}

\begin{figure}[h!]
    \centering
    \begin{minipage}{0.49\columnwidth}
        \centering
        \includegraphics[width=0.8\textwidth]{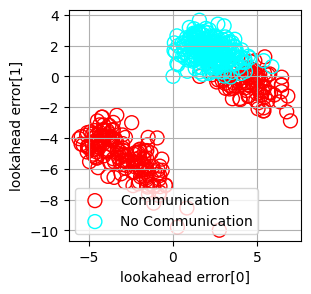} 
    \end{minipage}\hfill
    \begin{minipage}{0.49\columnwidth}
        \centering
        \includegraphics[width=0.8\textwidth]{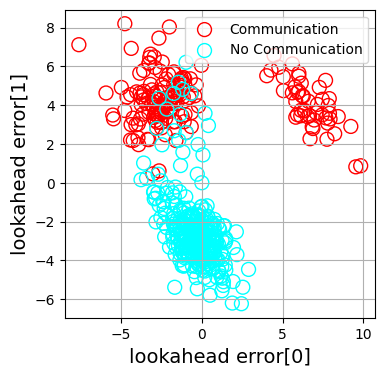} 
    \end{minipage}
    \caption{\small{Scheduling landscapes for a 2-mode GMM (left) and 3-mode GMM (right) for the inverted pendulum system (red denotes transmissions while cyan denotes silence).}}
    	\label{Fig:Linscatter_IP}
 	\vspace{-0.1cm}
\end{figure}

\begin{figure}[h!]
	\centering	\includegraphics[width=0.9\columnwidth ]{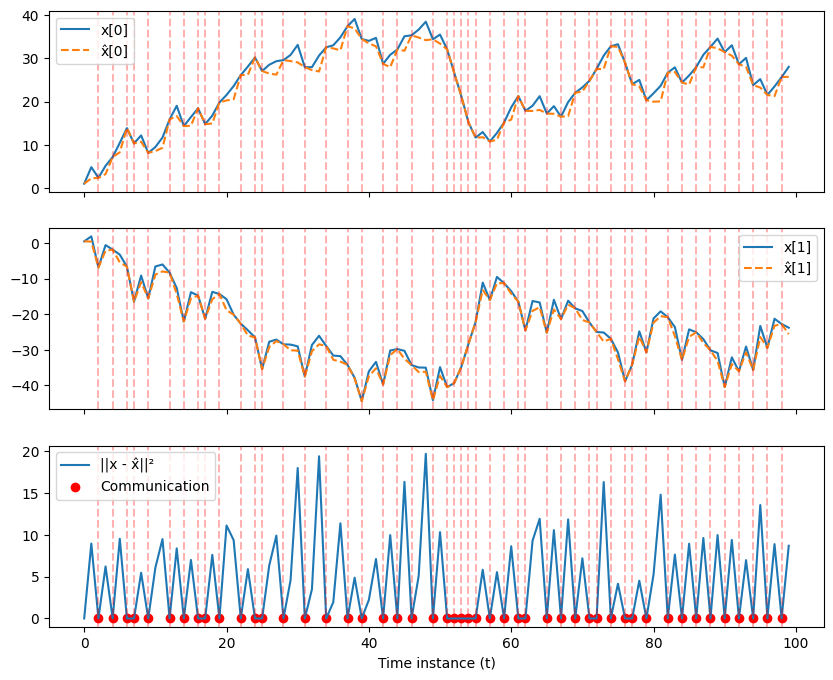}
 \vspace{-0.2cm}
	\caption{\small{Signal trajectories with communication instances (red markers) corresponding to (the left subfigure in) Fig.~\ref{Fig:Linscatter_IP} for the inverted pendulum system.
	}}
	\label{Fig:Lintraj_IP}
 	\vspace{-0.1cm}
\end{figure}

In Fig. \ref{Fig:traj_IP}, we plot the state, state estimate and the estimation error trajectories (overlayed with the communication instances) corresponding to the (left) scatter plot in Fig. \ref{Fig:scatter_IP} of a 2-mode GMM noise distribution. Additionally, in Fig. \ref{Fig:Linscatter_IP}, we also plot a comparative scatter plot obtained by training and evaluating a baseline linear estimator on a 2- and a 3-mode GMM noise distribution similar to the \texttt{CALM} algorithm results of Fig. \ref{Fig:scatter_IP}. The corresponding signal trajectories for the 2-mode GMM case are also plotted in FIg. \ref{Fig:Lintraj_IP}.

\subsection{Van der Pol}
\begin{figure}[h!]
	\centering	\includegraphics[width=0.9\columnwidth ]{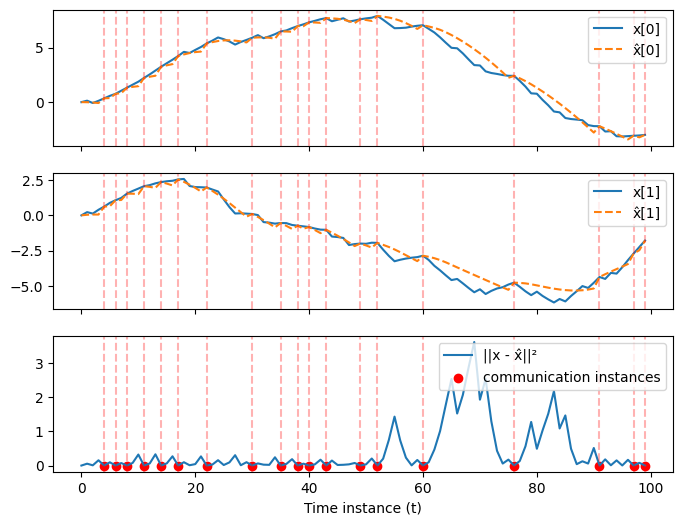}
 \vspace{-0.4cm}
	\caption{\small{Signal trajectories with communication instances (red markers) corresponding to the left subfigure in Fig.~\ref{Fig:scatter_vdp} ($\lambda = 0.7$) for the VdP oscillator. 
	}}
	\label{Fig:traj_vdp}
 	\vspace{-0.2cm}
\end{figure}

In Fig. \ref{Fig:traj_vdp}, we plot the state, state estimate and the estimation error trajectories, overlapped with the communication instances (red markers) corresponding to the scatter plot of the left subfigure in Fig. \ref{Fig:scatter_vdp}.

\subsection{Robot trajectory tracking}
\begin{figure*}
    \centering
    \begin{minipage}{0.5\textwidth}
        \centering
        \includegraphics[width=\textwidth]{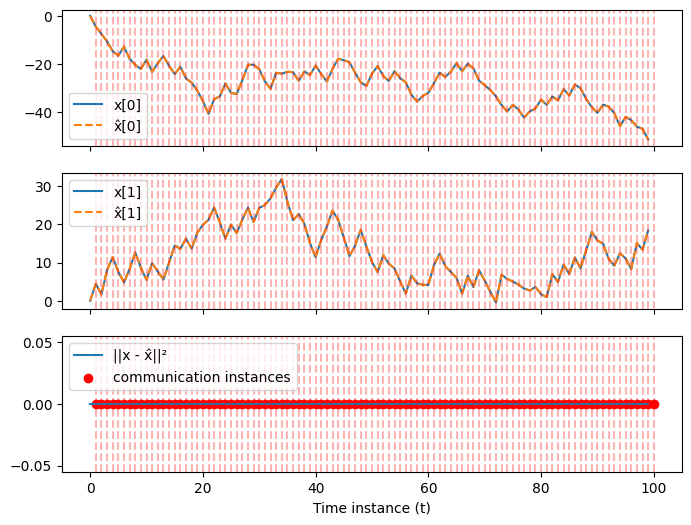} 
    \end{minipage}\hfill
    \begin{minipage}{0.5\textwidth}
        \centering
        \includegraphics[width=\textwidth]{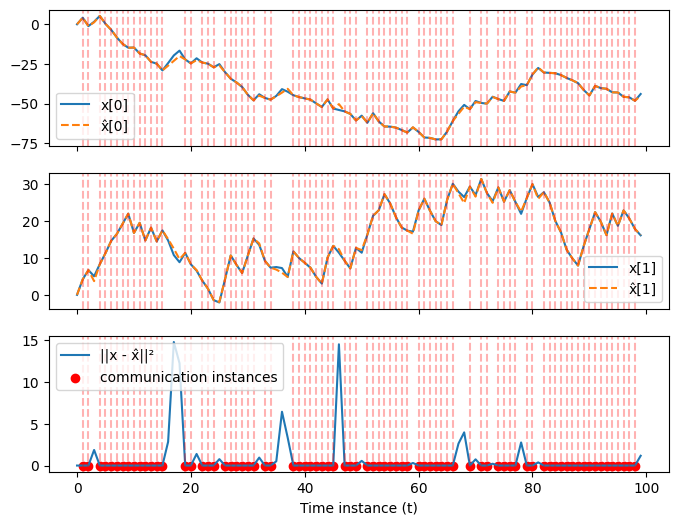} 
    \end{minipage}
    \begin{minipage}{0.5\textwidth}
        \centering
        \includegraphics[width=\textwidth]{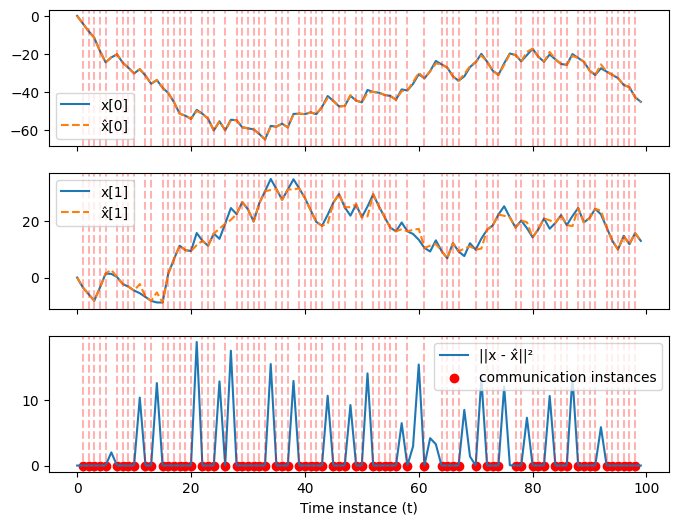} 
    \end{minipage}\hfill
    \begin{minipage}{0.5\textwidth}
        \centering
        \includegraphics[width=\textwidth]{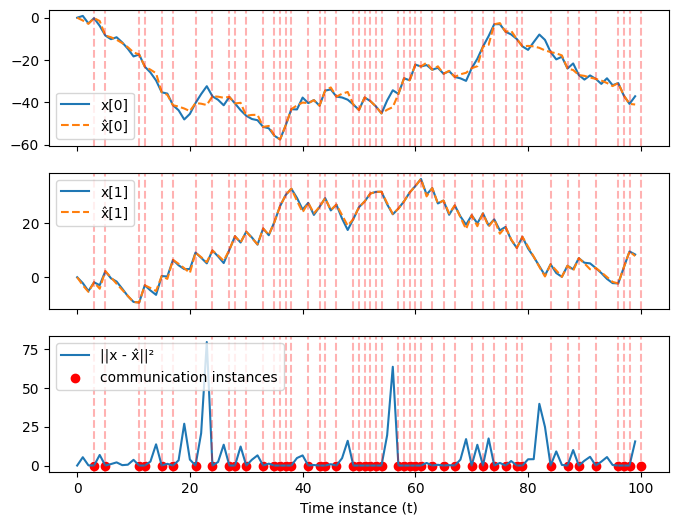} 
    \end{minipage}
    \caption{\small{Signal trajectories over time for the robot trajectory tracking system: $\lambda=15$ (top left), $\lambda=30$ (top right), $\lambda=40$ (bottom-left) and $\lambda=70$ (bottom right).}}
    	\label{Fig:traj_pt}
 	\vspace{-0.2cm}
\end{figure*}

For the trajectory tracking system, we plot the signal trajectories for each value of $\lambda$ corresponding to Fig. \ref{Fig:traj_pt}. As $\lambda$ increases from 15 to 30 to 40 to 70, the number of communication instances decrease from 499 to 386 to 374 to 249, as aligned with intuition.

\begin{figure}
	\centering	\includegraphics[width=0.9\columnwidth ]{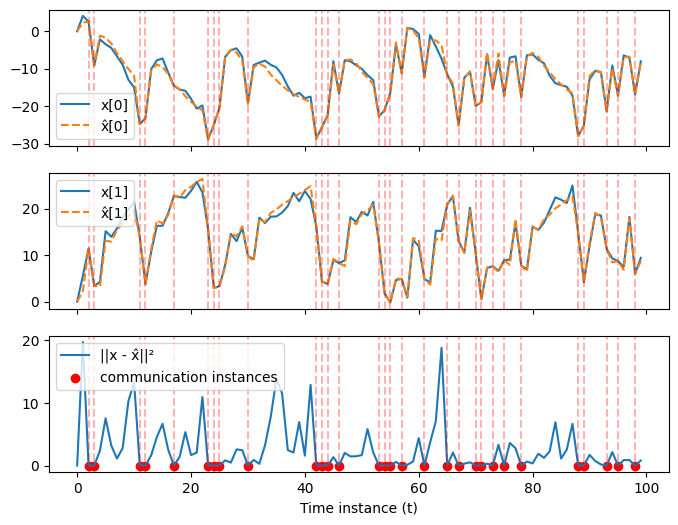}
 \vspace{-0.4cm}
	\caption{\small{Signal trajectories with communication instances (red markers) corresponding to Fig.~\ref{Fig:scatter_bo} for the Boeing flight control system.
	}}
	\label{Fig:traj_bo}
 	\vspace{-0.2cm}
\end{figure}

\end{document}